\documentclass[prd,twocolumn]{revtex4}
\usepackage{graphicx, epsfig}
\usepackage{color}
\usepackage{mathrsfs}
\usepackage{amsmath}

\newcommand{\be}{\begin{equation}}
\newcommand{\ee}{\end{equation}}
\newcommand{\bea}{\begin{eqnarray}}
\newcommand{\eea}{\end{eqnarray}}

\newcommand{\gapp}{\mathrel{\raise.3ex\hbox{$>$}\mkern-14mu
              \lower0.6ex\hbox{$\sim$}}}
\newcommand{\lapp}{\mathrel{\raise.3ex\hbox{$<$}\mkern-14mu
              \lower0.6ex\hbox{$\sim$}}}

\begin{document}
\title{Constraints on dark matter particles charged under a hidden gauge group from primordial black holes}
\author{De-Chang Dai$^1$, Katherine Freese$^2$ and Dejan Stojkovic$^1$}
\affiliation{$^1$HEPCOS, Department of Physics,
SUNY at Buffalo, Buffalo, NY 14260-1500}
\affiliation{$^2$MCTP, Department of Physics,
University of Michigan, Ann Arbor, MI 48109}


\begin{abstract}
\widetext
In order to accommodate increasingly tighter observational constraints on dark matter, several models have been proposed recently in which dark matter particles are charged under some hidden gauge group. Hidden gauge charges are invisible for the standard model particles, hence such scenarios are very difficult to constrain directly. However black holes are sensitive to all gauge charges, whether they belong to the standard model or not.  Here, we examine the constraints on the possible values of the dark matter particle mass and hidden gauge charge from the evolution of primordial black holes. We find that the existence of the primordial black holes with reasonable mass is incompatible with dark matter particles whose charge to mass ratio is of the order of one. For dark matter particles whose charge to mass ratio is much less than one, we are able to exclude only heavy dark matter in the mass range of $10^{11}$GeV-$10^{16}$GeV. Finally, for dark matter particles whose charge to mass ratio is much greater than one, there are no useful limits coming from primordial black holes.
\end{abstract}


\maketitle

\section{Introduction}

One of the outstanding problems in modern cosmology is the dark matter problem, or the problem of the missing mass in the universe.
Many solutions have been proposed so far, yet it looks like we are still far from the satisfactory resolution. The usual dark matter suspect is a particle with the weak interaction cross section and the mass around a TeV, since such a particle automatically satisfies the relic density constraints.  However, new observational data seem to indicate that a simple solution is unlike to satisfy all of the constraints. This was the motivation behind new proposals (see e.g. \cite{Ackerman:2008gi,Feng:2008ya,Hooper:2008im,Kim:2008pp,Huh:2007zw,Chun:2008by,Kaloper:2009nc,Feng:2009mn}) in which dark matter particles are charged under some hidden gauge group.

On the other hand, primordial black holes are a very important imprint of the physics in the early universe. When density fluctuations in some region are so high that the gravity can overcome the pressure, then the whole region collapses and makes a black hole. These density fluctuations can be caused by many different mechanisms, like oscillations of the inflaton field, and first and second order phase transitions \cite{Hawking:1971ei}. Collapse of the closed strings loops or closed domain walls also end in primordial black hole formation. While Cosmic Microwave Background provides very good constraint on the spectral index of perturbation on very large scales, so far primordial black holes are the only viable probe of the small scale spectral index of perturbations.

In the literature, there are studies trying to provide limits on the allowed mass fraction of primordial black holes from the requirement that these black holes do not change the standard cosmology \cite{MacGibbon:1991vc}. There are also studies using primordial black holes as a solution to some cosmological problems \cite{SF}. Decay of primordial black holes could even source gravitational waves \cite{Anantua:2008am}.
Here, we will try to connect the physics of primordial black holes with the dark matter charged under a hidden gauge group. According to no-hair theorems, a black hole state is uniquely characterized by its mass, angular momentum and gauge charges. If the initial state of the collapsing matter carries some gauge quantum number (not necessarily the standard model gauge charge), information about this quantum number can not be lost during the evolution of the black hole. Thus primordial black holes carry information about the composition of the cosmic fluid at early times.

Imagine that the dark matter is charged under a hidden gauge group, say $U(1)$. A black hole of an initial mass $M_i$ is formed predominantly of dark matter, so it contains approximately $N_0=M_i/m$ dark matter particles where $m$ is the dark matter particle mass. Most of these dark matter particles would annihilate with each other once inside the horizon, leaving the residual number of $N\sim \sqrt{N_0}$ particles with the charge of the same sign. If the charge of the individual particle is $q$, then the total charge of the black hole is $Q=\sqrt{N_0} q$.
If during its evolution a black hole can not get rid of all of its charge, then such a black hole can not evaporate completely. Instead, a stable remnant in the form of an extremal Reisner-Nordstrom black hole would be formed. An extremal black hole has its mass equal to its charge. Such a black hole has zero Hawking temperature and it does not evaporate any further. In order not to violate the observational constraints, the black hole relics  must not overclose the universe at the present time, i.e. the condition $\Omega_{pbh,0} <1$ must be fulfilled. Under certain conditions it is possible to form stable charged remnants which are not extremal, but can not decay further because of kinematical reasons, i.e. there are no available decay channels which satisfy both charge and energy conservations.

The condition that the black hole relics must not overclose the universe can be used to find the constraints on the mass and (hidden) gauge charge of dark matter. We are interested only in primordial black holes whose initial mass is less than $5 \times 10^{14}$g. Black holes heavier than this would not have decayed by now and can not provide any constraint on dark matter in our context. There are also independent constraints on the abundance of primordial black holes from the requirement that their evaporation should not affect nucleosynthesis, cosmic microwave background radiation, diffuse gamma ray background etc (see e.g. \cite{Green:1997sz}). However, in our context, the remnants will cut off evaporation and these constraints are not valid anymore.

\section{Basic assumptions}

We assume that primordial black holes are produced by density fluctuations due to oscillations of some (scalar)
field. If, within some region of space, density fluctuations
are large \cite{Green:1997sz,Kodama:1982sf,Crawford:1982yz,Hsu:1990fg,Sanz:1985np}, the whole region, usually the whole horizon volume at that time, collapses and forms a black hole. Since the horizon volume at some later time contains many earlier time horizon volumes, there are many black holes within a horizon at some later time.
The horizon mass in a radiation-dominated universe is given by
\begin{equation}
\label{eqn:Mass}
M_H \cong 10^{18}g \left(\frac{10^7 GeV}{T}\right)^2.
\end{equation}
where, $T$ is the temperature in the universe. Obviously, the earlier  the black holes are formed, the lighter they are, and the shorter their lifetime.

The main constraint applied in this paper is that the present mass density of today's relics of
primordial black holes must not overclose the universe, i.e.,
\begin{eqnarray}
\label{eqn:density-1}
\Omega_{rel,0}&=&\frac{M_{rel}}{M_{i}}\frac{\rho_{pbh,i}}{\rho_{total,0}}\left( \frac{T_0}{T_{form}}\right)^3\nonumber\\
&\sim &\frac{M_{rel}}{M_{i}}\beta_i \frac{\rho_{rad,i}}{\rho_{total,0}}\left( \frac{T_0}{T_{form}}\right)^3 <1.
\end{eqnarray}
where $\Omega_{rel,0}$ is the density parameter for black hole relics today, $\rho_{pbh,i}$ is the initial black hole mass density, while $\rho_{total,0}$ is the total mass density of matter today.  $T_0$ and $T_{form}$ are the temperatures in the universe today and at the time when the black holes were formed respectively. The parameter $\beta_i$ is the mass fraction of the universe in black holes when they were formed, with $\rho_{rad,i}$ being the radiation energy density at that time.

In general, black holes with mass $M_{i} < 5\times 10^{14}$g decay via Hawking evaporation before the current time. Motivated by the information loss paradox \cite{Hawking:1976ra,Page:1993up,Vachaspati:2006ki,Vachaspati:2007hr,Greenwood:2008ht}, some authors argued that black holes may not evaporate completely \cite{Bowick:1988xh,Coleman:1991sj,Gibbons:1987ps,Garfinkle:1990qj,Preskill:1991tb,Hossenfelder:2001dn} and instead leave a Planck mass relics, i.e. $M_{rel}\sim M_{Pl}$. If this is true, then a primordial black hole with the initial mass $M_{i}< 5 \times 10^{14}g$ will become a relic with the mass $M_{Pl}$.

Though the possibility of a neutral black hole leaving a remnant seems unlikely, once we include the gauge charges the situation becomes more complicated. At the moment a primordial black hole is made, it already contains a non-zero net charge as explained in the Introduction. The net quantum number must be released in order for the black hole to evaporate completely. This is true for all gauge charges. If a black hole can not get rid of its gauge charge, at a certain moment it can become (close to) extremal, and its Hawking temperature will drop close to zero. Such black holes can have a relic mass larger than $M_{Pl}$.

Whether it is possible to for a black hole to evolve into an  extremal black hole depends on the parameters in theory. To cover all the possibilities, we consider here three different ratios between the charge and mass of dark matter.

\begin{itemize}

\item[(1)] $q/m \gg 1$

\item[(2)] $q/m \ll 1$

\item[(3)] $q/m \sim 1$

\end{itemize}
First we consider the case $(1)$ where the dark matter charge is much larger than its mass. The familiar analog is an electron, since its charge is much larger than its mass in Planck units, i.e. $e/m_e = (\sqrt{\alpha}M_{Pl})/(10^{-21}M_{Pl}) \approx 10^{20}$. For charged black holes, greybody factors depend not only on the mass of the black hole and emitted particles but also on their charges \cite{Dai:2007ki}. It is not only the pure Hawking effect that causes the emission of particles, but also the electromagnetic energy of the black hole. Electric field of the black hole induces pair creation in vacuum. Since electron charge is much greater than its mass, the electromagnetic contribution to the electron greybody factor will be dominant over the mass contribution. A charged black hole strongly prefers to emit particles of the same sign since they penetrate the potential barrier easier. Therefore, the black hole can discharge its electric charge easily. Its electric charge will go to zero much faster than its mass. Once that happens, the black hole continues its evaporation as an ordinary Schwarzschild black hole. Therefore, the case $(1)$, which is an analog of the electron case, can not leave stable relics which may endanger standard cosmology.  The only exception may be if primordial black holes are formed already very close to extremal. While there exists the possibility that such black holes can survive until today it is hard to image that most of the primordial black holes were formed extremal. Therefore, the case $(1)$ can not provide many relic black holes.

The case $(2)$ is the one in which the dark matter mass is much larger than its charge. In this case, a black hole can not get rid of its charged particles as easily as in the case $(1)$. The charge contribution to the greybody factors is much smaller in this case, so the mass contribution will be dominant.  Since heavy particles can be emitted only when the black hole temperature is higher than the particle mass, the black hole will lose most of its mass by emitting other lighter particles that do not carry the quantum number in question. At some later stage of evaporation, the black hole mass is just a fraction of its original mass, but its charge is practically the same as the original. In order to get rid of $N$ units of charge, the black hole needs to emit $N$ dark matter particles, but it does not have available mass to achieve this. At the last stage of evaporation, when the remaining mass of the black hole is of the order of $M_{Pl}$, the final burst of particles is expected to take place. Since the black hole has to respect the conservation of the gauge charge \cite{Stojkovic:2005zq}, as a result, a stable charged remnant of about $M_{Pl}$ must be formed. Note that this is not an extremal black hole, but it is a remnant which can not decay further for kinematical reason, i.e. there are no available decay channels that obey both charge conservation and kinematical requirements.
If the temperature of a black hole is always greater than the mass of the dark matter particle (which can happen for a light primordial black hole or a light dark matter particle), then such a black hole
is unlikely to leave remnants.
Therefore, the crucial requirement for the remnants to form is that the temperature of the black hole is smaller than the mass of the dark matter particle. Then, such a black hole loses its mass by emitting other particles and, though at some point may emit a few dark matter particles it will never be able to get rid of all of its charge. We will quantify this requirement in Section~\ref{constraints}.

\begin{figure}[t]
   \centering
\includegraphics[width=3.35in]{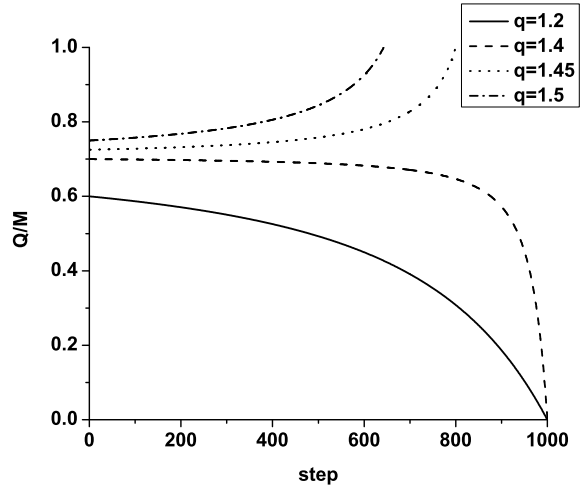}
\caption{Evolution of the charge to mass ratio of the black hole, $Q/M$. In this figure we assume that a black hole emits only dark matter particles with $q/m \sim 1$ (case $3$). The total initial mass of the black hole is $2000 M_{Pl}$ while its charge is $Q=1000q$. We make plots for several different values of the dark matter charge $q$ in Planck units: $q= 1.2$, $q= 1.4$, $q = 1.45$ and $q=1.5$. We find that a black hole always become extremal, i.e. $Q/M$ always evolves to $1$, if the initial $Q/M \gtrsim 0.725$ (corresponding to $q>1.45$ in the figure). }
\label{fig:q-m}
\end{figure}

The case $(3)$ is the one in which dark matter mass is comparable to its charge. This case is between the cases $(1)$ and $(2)$, and allows for the extremal black hole relics whose mass is much larger than $M_{Pl}$. A simple estimate of the evolution of the charge to mass ratio of the black hole, $Q/M$, is shown in Fig.~\ref{fig:q-m}. A black hole is assumed to emit dark matter particles only with $q/m \sim 1$. The total initial mass of the black hole is $2000 M_{Pl}$ while its charge is $Q=1000q$; i.e. in this case $Q/M = q/(2 M_{pl})$. Note that the total mass/energy of the black hole does not need to be equal to the sum of the masses of its constituencies because of the potential energy of interaction between $Q$ and $q$. In each step this black hole loses a portion of its mass $\Delta M$ which is roughly equal to the mass of the emitted particle plus the potential energy of interaction (which is this case can not be neglected)
\be \Delta M = m +Qq/R_h
\ee
where $R_h$ is the horizon radius of the charged black hole in Planck units
\be
R_h = M + \sqrt{M^2 - Q^2} \, .
\ee
The question is whether $Q/M$ of the black hole can evolve to $1$ before the black hole loses all of its energy. We make plots for several different values of the dark matter charge $q$. We find that a black hole always becomes extremal, i.e. $Q/M$ always evolves to $1$, if the initial $Q/M \gtrsim 0.725$. Since the dark matter is much more abundant than ordinary matter, the condition $Q/M \gtrsim 0.725$ is easy to satisfy if a black hole was formed mostly of dark matter particles. Thus, if $q/m \sim 1$, most of the primordial black holes end up being extremal black holes.

From the phenomenological point of view, one should be careful in dealing with particles with $q/m \sim 1$. Such particles (with the same sign of charge) would feel gauge repulsion in addition to gravitational attraction, so they may not cluster easily. If these charged particles dominate over the uncharged ones, that would make structure formation difficult. However, note that our estimate is very conservative since we ignored many standard model particle decay channels which take away the black hole mass while leaving its hidden gauge charge unaffected. Thus, until the temperature of the black hole reaches the value at which it can emit dark matter particles (say a TeV), $Q/M$ of the black hole will go up. Thus, a black hole does not have to be formed dominantly of charged particles in order to reach the critical point at which it becomes extremal.

To reiterate, case $(1)$ does not produce many relic black holes;  case $(2)$ can leave relic black holes
due to kinematic constraints if the black hole temperature at some (early) point in its evolution is smaller than the dark matter particle mass; and case $(3)$ generically produces extremal black hole relics.

\section{Constraints}
\label{constraints}

Since the case $(3)$ looks the most promising, we will first focus on it. Consider a black hole with the net gauge charge $Q= Nq$ which became  extremal. For an order of magnitude estimate, we assume that the black hole became extremal with a charge which is not much different from the initial net charge at the formation (Fig.~\ref{fig:q-m} supports this assumption). The mass of this extremal relic is
\begin{equation}
\label{eqn:Mrel}
M_{rel}\sim Q =Nq \sim Nm.
\end{equation}
If a black hole is formed in a collapse of the total number of $N_0$ dark matter particles, we can expect the net charge of the newly formed black hole to be only $N \sim \sqrt{N_0}$, since the total number $N_0$ counts both particles and antiparticles, many of which would annihilate with each other so that only a residual random walk charge remains. Thus,
\begin{eqnarray}
\label{eqn:N}
N &\sim &\sqrt{\frac{M_{i}}{m} } \mbox{ if  } m>T_{form}\nonumber\\
  &\sim &\sqrt{\frac{M_{i}}{T_{form}} } \mbox{ if  } m<T_{form}
\end{eqnarray}
Combining this equation with Eq.~(\ref{eqn:Mrel}), we can find the mass of the relic black hole. The initial black hole mass $M_{i} \sim N_0 m$ is determined by the mass within the horizon at the time of formation, which is given by Eq.~(\ref{eqn:Mass}). In order not to overclose the universe, Eq.~(\ref{eqn:density-1}), the mass fraction of the black holes has to satisfy (see \cite{Green:1997sz})

\begin{equation}
\label{eqn:b1}
\beta _i<1.5\times 10^{-29}\frac{M_{i}^{3/2}}{M_{Pl}^{1/2}M_{rel}}
\end{equation}

The mass fraction $\beta_{i}$ can be found assuming a Gaussian perturbation spectrum of fluctuations that leads to  black hole formation with (see \cite{Green:1997sz,Liddle:1993fq,Bringmann:2001yp})
\begin{equation}
\label{eqn:b2}
\beta_i \approx \frac{\sigma_H (T_{form})}{\sqrt{2\pi}\delta_{min}}e^{-\frac{\delta_{min}^2}{2\sigma^2_H (T_{form})}}
\end{equation}
where $T_{form}$ is the temperature when the primordial black holes were formed, while $T_0$ is the temperature today. The parameter $\delta_{min}\approx 0.3$ is the minimum density contrast required to form a primordial black hole. $\sigma_H (T)$ is the mass variance evaluated at horizon crossing at the temperature $T$. It is defined as \cite{Liddle:1993fq}
\begin{equation}
\label{eqn:fluc}
\sigma_H (T_{form})=\sigma_H (T_0) \left( \frac{M_H (T_0)}{M_H (T_{eq})}\right)^{\frac{n-1}{6}}\left( \frac{M_H(T_{eq})}{M_H(T_{form})}\right) ^{\frac{n-1}{4}}
\end{equation}
where $T_{eq}$ is the temperature at the matter/radiation equilibrium, while
$n$ is the spectral index of scalar fluctuations that lead to the black hole formation, i.e. $P(k)\propto k^n$. Recent WMAP data indicate  that the value of $n$ of scalar fluctuations of the inflaton field is near the Harrison-Zel'dovich type, $n\approx 1$. However WMAP data probe the scales between $10^{45}$ and $10^{60}$ times larger than those probed by primordial black holes, which can be formed by fluctuations of fields different from the inflaton (e.g. during phase transitions). The value of $n$ frequently used in this context is between $1.23$ and $1.31$ \cite{Green:1997sz,Bringmann:2001yp,Kim:1999iv}. To normalize Eq.~(\ref{eqn:fluc}) we use the mass variance evaluated at the horizon crossing
\begin{equation} \label{norm}
\sigma_H(T_0)=9.5\times 10^{-5} \, .
\end{equation}

\begin{figure}[t]
   \centering
\includegraphics[width=3.35in]{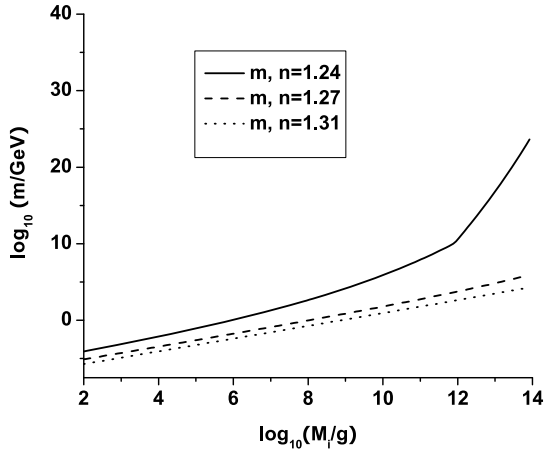}
\caption{Constraints on the values of the dark matter particle mass with $q/m \sim 1$ as a function of
black hole mass $M$ (in grams $g)$. The values above the curves $n=1.24$, $n= 1.27$ and $n= 1.31$ are forbidden. }
\label{fig:constraint}
\end{figure}

Fig.~\ref{fig:constraint} shows the constraint on the possible mass of dark matter particles coming from an eventual existence of primordial black holes of a given mass. We now describe the details of the procedure to make such a plot. We first choose some value of the initial black hole mass $M_i$. Eq.~(\ref{eqn:Mass}) determines the value of the temperature $T_{form}$ at which the black holes of that mass are formed. Eq.~(\ref{eqn:N}) gives us the value of $N$ as a function of the dark matter particle mass $m$. Then Eq.~(\ref{eqn:Mrel}) gives us the value of $M_{rel}$, again as a function of the dark matter particle mass $m$.
The main constraint is Eq.~(\ref{eqn:b1}), so we substitute Eq.~(\ref{eqn:b2}), along with Eqns. (\ref{eqn:fluc}) and (\ref{norm}), into the left-hand side of Eq.~(\ref{eqn:b1}). We then check if a given value of $m$ satisfies the constraint. We repeat the procedure for different values of $M_i$ and make the plot in Fig.~\ref{fig:constraint} for several values of the spectral index $n$. The values of $m$ above the curves $n=1.24$, $n= 1.27$ and $n= 1.31$ are forbidden. It is clear that the constraint is stronger for larger values of the spectral index $n$. For example, the spectral index of $n= 1.31$ excludes a large region in parameter space of the allowed values for the dark matter particles for reasonable masses of the primordial black holes. Also, if the primordial black holes are created earlier, with masses up to $10^{10}$g, which is the most likely scenario, the allowed mass of the dark matter particles is very small, much smaller than  $100$GeV. These results indicate that dark matter particles with $q/m \sim 1$ are barely compatible with the existence of the primordial black holes.

We now consider the case $(2)$ in which the dark matter charge is much smaller than its mass, ie. $q/m \ll 1$. As explained above, in this case stable charged (non-extremal) relics of mass $M_{Pl}$ can be formed.  The crucial requirement for relics to form is that the black hole temperature is lower than the dark matter particle mass, i.e.
\be \label{tbh}
T_{bh}=\frac{1}{4\pi M} <m
\ee
where the relevant mass of the black hole is
\be \label{M}
M =N m
\ee
where  $N \sim \sqrt{N_0}$ is the net charge of the black hole, with $N_0$ being the total number of dark matter particles that made the initial black hole of mass $M_i =N_0 m$.
If condition in Eq.~(\ref{tbh}) is fulfilled, a (non-extremal) relic of mass $M_{Pl}$ will form which can not decay further for kinematical reasons. Then, we can apply similar analysis as we did for the case $(3)$. The difference is that now $M_{rel}=M_{Pl}$. Also, for any given point $(M_i,m)$ on the plot, we check if the condition in Eq.~(\ref{tbh}) is fulfilled. We do that by finding $N$ from Eq.~(\ref{eqn:N}) for a given pair $(M_i,m)$, which we then substitute into Eq.~(\ref{M}) to get the value for $M$. This $M$, substituted into Eq.~(\ref{tbh}), decides if the relic of mass $M_{Pl}$ will be formed for a given pair $(M_i,m)$. If the answer is positive, Eq.~(\ref{eqn:b1})  gives the constraint on the dark matter particle mass $m$ as a function of the initial primordial black hole mass $M_i$. We note that the constraint in Eq.~(\ref{tbh}) is stronger that the constraint in Eq.~(\ref{eqn:b1}), so if a pair $(M_i,m)$ satisfies the former it is practically guarantied that it will satisfy the latter constraint.
\begin{figure}[t]
   \centering
\includegraphics[width=3.35in]{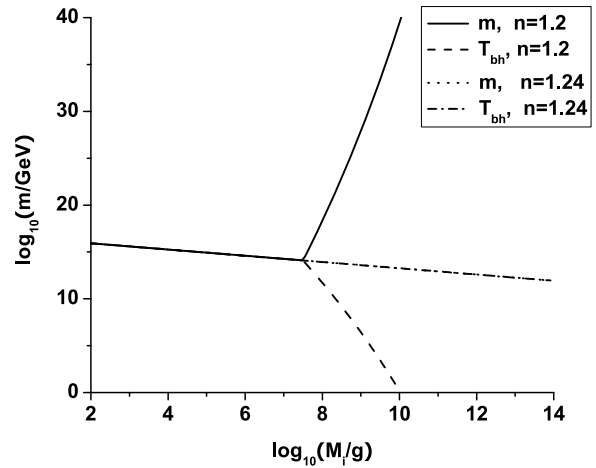}
\caption{Constraints on the values of the dark matter particle mass with $q/m \ll 1$. The values above the curves labeled ($m, n=1.2$) and ($m, n=1.24$) are forbidden. The black hole temperature $T_{bh}$ from Eq. (\ref{tbh}) is plotted simultaneously with $m$ for convenience. }
\label{fig:q-small}
\end{figure}

Fig.~\ref{fig:q-small} shows the constraint on the dark matter particle mass with $q/m \ll 1$, for two values of the spectral index $n$. The constraint is much weaker than that in the case $(3)$ plotted in Fig.~\ref{fig:constraint}. Here, in case $(2)$, the mass of the dark matter particle or the mass of the initial primordial black hole mass has to be quite large in order to overclose the universe. For example, we can conclude that the existence of the primordial black holes in the mass range of $10^2$g-$10^{15}$g is only incompatible with  dark matter particles with extremely high masses in the range of  $10^{11}$GeV-$10^{16}$GeV (lighter black holes exclude heavier dark matter and vice versa), for the spectral index $n=1.24$. Higher values of the spectral index make the constraints stronger.

\section{Conclusion}

Beyond the standard model theories, motivated by potential solutions to some fundamental problems in particle physics and cosmology, often include particles charged under some hidden gauge group.
Since hidden gauge charges are invisible for the standard model particles, such models are very difficult to constrain directly. As it is well known, black holes do not distinguish between the standard model and hidden gauge charges, and can provide important constraints on such models. We explored here the constraints on the possible values of the dark matter particle mass and hidden gauge charge from the evolution of primordial black holes. We assumed that primordial black holes are made in the early universe mostly of dark matter particles. Depending on the the particular vales of the mass and charge of these particles, a primordial black hole can either evaporate completely or leave a stable remnant. Stable remnants are potentially dangerous since they can overclose the universe.

We showed that if the charge to mass ratio of the dark matter particle is of the order one, a primordial black hole generically becomes extremal (see Fig.~\ref{fig:q-m}). We then calculated the mass of the dark matter particle which would lead to the overclosure the universe.
For generic parameters, we showed that practically all of the reasonable values for the dark matter mass are excluded (see Fig.~\ref{fig:constraint}).
This implies that the existence of the primordial black holes created in very early universe is incompatible with dark matter particles whose mass is of the same order of magnitude as its charge.

In the case when  the dark matter mass is much larger than its charge, the electromagnetic contribution to the greybody factors is much smaller than the mass contribution. Since heavy particles can be emitted only when the black hole temperature is higher than the particle mass, the black hole will lose most of its mass by emitting other lighter particles that do not carry the quantum number in question. Since the black hole has to respect the conservation of the gauge charge, as a result, a stable charged (non-extremal) remnant of about $M_{Pl}$ must be formed. This remnant can not decay further for kinematical reason, i.e. there are no available decay channels that obey both charge conservation and kinematical requirements. Again, the requirement that the remnants do not overclose the universe puts constraints on dark matter mass. For example, from Fig.~\ref{fig:q-small} we see that the existence of the primordial black holes in the mass range of $10^2$g-$10^{15}$g is incompatible with dark matter in the range of  $10^{11}$GeV-$10^{16}$GeV for the spectral index of $n=1.24$ (this is a very short scale spectral index of perturbations during the formation of the primordial black holes, and not the one constrained by CMB fluctuations to be $n\approx 1$). In this regime, when the the dark matter mass is much larger than its charge, lighter black holes formed earlier exclude heavier dark matter and vice versa. Higher values of the spectral index, as expected, make the constraints stronger.

Finally, in the case when the the dark matter charge is much larger than its mass, the evolution of the primordial black holes is very unlikely to leave stable remnants. The reason is that the charge contribution to the greybody factors is much greater than the mass contribution, and the black hole can easily get rid of its charged particles. In this regime, primordial black holes can not provide any useful constraint on dark matter mass.

Obviously, our constraints are sensitive to the primordial black hole abundance. Without primordial black holes, our constraints do not work. Generically, one expects large black hole production in small field inflation models (as opposed to the large field chaotic inflation models where no significant amount of black holes are produced), and during the phase transitions in our universe.

We comment here on how the recent conjecture that
gravity should always be the weakest force \cite{ArkaniHamed:2006dz} may affect our results.
If the conjecture in \cite{ArkaniHamed:2006dz} is correct, then the mass of the remnant might not be determined fundamentally only by the scale of gravity $M_{Pl}$, but also by the strength of the gauge coupling $\alpha$.
If this is true, then instead of the remnants of the order of Planck mass, one might have remnants of the mass $\sqrt{\alpha} M_{Pl}$, which is lighter than $M_{Pl}$ for small $\alpha$. Currently, the observed halo structure of dark matter in galaxies places an upper
bound of $\alpha < 10^{-3}$ \cite{Ackerman:2008gi}. Then, our constraints may be avoided by taking arbitrarily small $\alpha$. This also opens a new interesting possibility that remnants themselves can play the role of dark matter, since e.g. $\alpha = 10^{-30}$ gives the mass of the remnant of about a TeV.

\begin{acknowledgments} DD and DS are grateful to the organizers of the MCTP workshop ``Dark Side of the Universe II" at the University of Michigan for hospitality. We thank N. Kaloper and D. Spolyar for very useful conversations. DS acknowledges the financial support from NSF. KF thanks the MCTP and the DOE for support via
the University of Michigan.
\end{acknowledgments}

\end{document}